**TITLE:** **A gel wax phantom for performance evaluation in diagnostic ultrasound: assessment of image uniformity, geometric accuracy, and diameter of a hyperechoic target**

**AUTHOR NAMES AND AFFILIATIONS**

1. **Ms. Debjani Phani** (Corresponding Author)

   **Address for Communication:**

   Assistant Professor, Department of Radiation Physics, Regional Cancer Centre, Thiruvananthapuram- 695011, State: Kerala, INDIA.

   **E-mail Address:** debjaniphani@gmail.com

   **ORCID ID 0000-0003-0171-4411**

   **Affiliations:**

   a) Department of Radiation Physics, Regional Cancer Centre, Thiruvananthapuram-695011, Kerala, INDIA.

   b) Meenakshi Academy of Higher Education and Research, Chennai-600 078, Tamil Nadu, INDIA

2. **Prof. Rajasekhar Konduru Varadarajulu**
   **Affiliation:** c) Meenakshi Medical College Hospital & Research Institute, Chennai - 631552, Tamil Nadu, INDIA.

   E-mail Address: roshoraju@gmail.com

3. **Arijit Paramanick**
   **Affiliation:** d) School of Physics, Indian Institute of Science Education and Research Thiruvananthapuram (IISER-TVM), Vithura, Thiruvananthapuram, Kerala 695551, India

   E-mail Address: arijit.paramanick21@iisertvm.ac.in

   **ORCID ID 0000-0002-7833-3337**

4. **Souradip Paul**
   **Affiliation:** d) School of Physics, Indian Institute of Science Education and Research Thiruvananthapuram (IISER-TVM), Vithura, Thiruvananthapuram, Kerala 695551, India

E-mail Address: souradip.rkm16@iisertvm.ac.in

**ORCID ID 0000-0003-1605-6087**

**5.    Dr. Raghukumar Paramu**

**Affiliation:**    a) Department of Radiation Physics, Regional Cancer Centre, Thiruvananthapuram- 695011, Kerala, INDIA.

E-mail Address: raghu.rcc@gmail.com

**6.    Mr. George Zacharia**

**Affiliation:**    a) Department of Radiation Physics, Regional Cancer Centre, Thiruvananthapuram- 695011, Kerala, INDIA.

E-mail Address: george.zcharia@gmail.com

**7.    Dr. Shaiju VS**

**Affiliation:**    a) Department of Radiation Physics, Regional Cancer Centre, Thiruvananthapuram- 695011, Kerala, INDIA.

E-mail Address: shaijuvs@gmail.com

**ORCID ID: 0000-0002-6390-0669**

**8.    Prof. Venugopal Muraleedharan**
**Affiliation:** d) Department of Radio Diagnosis, Regional Cancer Centre,

Thiruvananthapuram- 695011, Kerala, INDIA.

E-mail Address: drvenurcc@gmail.com

**9.    Dr. M. Suheshkumar Singh**

**Affiliation:**    d) School of Physics, Indian Institute of Science Education and Research Thiruvananthapuram (IISER-TVM), Vithura,

Thiruvananthapuram, Kerala 695551, India

E-mail Address: suhesh.kumar@iisertvm.ac.in

**10.    Dr. Raghuram Kesavan Nair**
**Affiliation:**    a) Professor (Retired), Department of Radiation Physics, Regional Cancer Centre, Thiruvananthapuram- 695011, Kerala, INDIA.
E-mail Address: raghurkn@gmail.com

ACKNOWLEDGMENT

The authors are grateful to Ramesh Babu V, Precision Fabrication Facility, Biomedical Technology Wing, Sree Chitra Tirunal Institute for Medical Sciences and Technology Trivandrum, Kerala, INDIA, for fabricating the polytetrafluoroethylene frames used in this study. We thank Prof. Jyoti R Seth, Advanced Rheology Facility, Indian Institute of Technology Bombay Powai, Mumbai, INDIA, for providing the information on polyethene gels and gel waxes.

**Ethical Statements**

**Ethical approval:**

This article does not contain any studies performed on human participants or animals.

**Conflict of interest:**

The authors declare that they have no conflict of interest.

**Funding**

The authors have no relevant financial or non-financial interests to disclose.

# Title: **A gel wax phantom for performance evaluation in diagnostic ultrasound: assessment of image uniformity, geometric accuracy, and diameter of a hyperechoic target**

## Abstract

**Purpose:** To develop and validate a phantom for diagnostic ultrasound (US) scanners by embedding targets in gel wax to determine the image uniformity, lateral and axial resolution, and diameter of a stainless-steel disc.

**Materials and Methods:** Acoustic property (AP), which includes the velocity of US ($c_{us}$), acoustic impedance (Z), and attenuation coefficient (μ) in gel wax were determined. The $c_{us}$, and μ were estimated using the pulse-echo technique. Z was obtained from the product of sample density (ρ) and $c_{us}$. Two rectangular frames using polytetrafluoroethylene (PTFE) sheets with holes separated by 5, 10, and 20 mm distances were constructed. Nylon filaments and SS-disc (diameter = 16.8 mm) were threaded through the frames and suitably placed in melted gel wax to obtain orthogonal targets. The targets were measured using computerized tomography (CT) and 2-9 MHz US probe.

**Results:** The AP of gel wax were $c_{us}$=1418 m/s, ρ= 0.87 g/cm$^3$, Z=1.23 MRayls, μ= 0.88 dB/cm/MHz. The results of US imaging of the targets were compared with their physical sizes and served as baselines for the scanner and probe. The maximum error in distance measurement in the phantom US images was 7.1%, and the phantom volume decreased by 1.8% over 62 weeks.

**Conclusion:** Gel wax can be useful in developing affordable, highly stable, and customizable diagnostic US phantoms that can be implemented widely.

# 1. Introduction

Many organizations have recommended quality assurance (QA) testing procedures for diagnostic ultrasound (US) equipment [1,2]. It is possible to identify the degradation in the image quality by testing every set of transducers. Adopting corrective measures at this stage will ensure safe and reliable patient scans. Although investigators have reported equipment faults in transducers, image formation, and physical flaws during US inspection, a routine QA is often disregarded [3,4]. The primary cause is the additional financial burden associated with the infrastructure required to perform these tests, which is rarely a regulatory requirement. US phantoms are essential to perform QA in US scanners. Commercial US QA phantoms are costly, and a single phantom cannot be used for all the tests that require different artifacts with varying echogenicity and geometry embedded in a tissue-mimicking material (TMM). This study attempts to address this problem by developing a low-cost phantom for the routine assessment of diagnostic US scanners. An US phantom should transmit US signals to image its artifacts, and be stable under ambient room conditions. Various materials such as agar, gelatin, and psyllium husk have been used as TMMs for clinical training in the diagnostic US, but they all have a limited shelf life [5–7]. *Madsen et al.* developed a milk-agar phantom that can last up to 2.5 years [8]. While these materials are suitable for designing training phantoms, QA phantoms should be mechanically stable and have a long shelf life (~ 5–10 years) to ensure reproducible imaging in routine practice.

Investigators have developed a material for US imaging applications using hydrocarbon monomers styrene-ethylene/butylene-styrene (SEBS), mixed with mineral oil where the resin allows the oil to set into a soft gel [9,10]. This transparent compound, containing 2–6% polymer resin and mineral oil, has been used to construct anatomical US phantoms, is known as copolymer-in-oil or gel wax [11–13]. It is evident from these studies that the composition of resin in mineral oil is different among different samples of gel wax and is unknown unless specified by the manufacturer. The acoustic properties (AP) of gel wax have been explored and it was mixed with additives such as paraffin wax, graphite powder, and glass microspheres to develop US phantoms mimicking human organs [1,14,15]. Gel wax was mixed with additives such as paraffin wax, graphite powder, and glass microspheres to vary its image contrast during US imaging and tune the AP [9,11]. The AP of a material is defined by a set of physical characteristics; the most critical are the velocity of US ($c_{us}$), acoustic impedance ($Z$), attenuation coefficient (µ), backscattering coefficient, and nonlinearity parameter [14]. An ideal TMM should have a $c_{us}$ equal to 1561 m/s (± 10 m/s), $Z$ of $1.6 \times 10^6$ Rayls, µ in the range of 0.5–0.7 dB/cm/MHz, and linear response of attenuation to frequency from 2 to15 MHz [1,14,15]. A material with similar properties will mimic the resolution, contrast, and penetration depth of soft tissue during US imaging. Our objective was to examine the possibility of using gel wax to develop a QA phantom for diagnostic US scanners. This work consisted of determining the AP of the gel wax, constructing the target elements, and embedding them into the gel wax to obtain the phantom. To ensure that the desired level of accuracy was maintained through all stages of testing and to document the imaging parameters phantom validation was performed. Finally, we estimated the change in phantom volume over time.

The AP of the gel wax was determined using the pulse-echo technique implemented in our previous study [16]. Target elements were incorporated in the gel wax to perform a subset of tests, as explained in the report *AAPM Ultrasound Task Group No. 1* for periodic evaluation of US scanners [1]. The present phantom design allows the evaluation of image uniformity, geometric conformity (distance accuracy) in orthogonal planes, and measurement of the diameter of a stainless steel (SS) disc (hyperechoic target) in a diagnostic US scanner. Considering the physical dimensions of each test element of the phantom as a reference, the corresponding variations in the computed tomography (CT) and US images were obtained. In addition, the US imaging provides the baseline values for the current phantom, scanner, and transducer set, which provides the peak performance of a scanner for a specific image quality indicator.

**2. Materials and methods**

The experimental setup and methodology to determine the AP of the gel wax are explained, followed by the steps to fabricate the target elements and the gel wax phantom. The CT and US images of the target embedded in the phantom were compared with the physical dimensions of the target. The stability of the phantom volume was also examined.

## *2.1 Determination of the acoustic property of gel wax*

### *2.1.1 Estimation of the velocity of ultrasound ($c_{us}$)*

Figure 1-a illustrates the experimental setup used to measure $c_{us}$ and µ. It consisted of a pulser-receiver (5073PR-40-P, Olympus, with a pulse repetition frequency of 200 Hz, a nominal frequency of 30 MHz, and a pulse width of nanoseconds), a focusing US transducer (Olympus V-375-SU with a 19 mm focal length, 153 nm focal spot size, and 4 mm focal zone length), a 3-Axis motorized positioning system (Newmark NSC-G Series, Newmark Systems Inc.), and a data acquisition (DAQ) system (779745-02, NI PCI-5114), as shown in Figure 1-b. The time-sharing feature allowed the pulser-receiver to detect the electrical signals resulting from the reflected acoustic pulses and enabled the amplification of weak signals (amplification ~ 39 dB). Samples were mounted (only two sides) on an acoustic reflector (an SS plate) fully immersed in a water bath (20 °C) with the transducer surface for acoustic coupling. To deliver maximum acoustic energy, the samples were kept at the focal zone of the transducer, and a train of US pulses was sent. After amplification, at a sampling frequency of approximately 100 MHz, the DAQ system received acoustic echoes (A-line data) reflected from the sample surfaces. The $c_{us}$ of the gel wax was determined from samples with three different thicknesses ($20 \times 20 \times 1.7$ mm$^3$, $20 \times 20 \times 2.2$ mm$^3$, and $20 \times 20 \times 2.6$ mm$^3$) that were prepared by melting the gel wax in a microwave oven and pouring into prefabricated plastic molds.

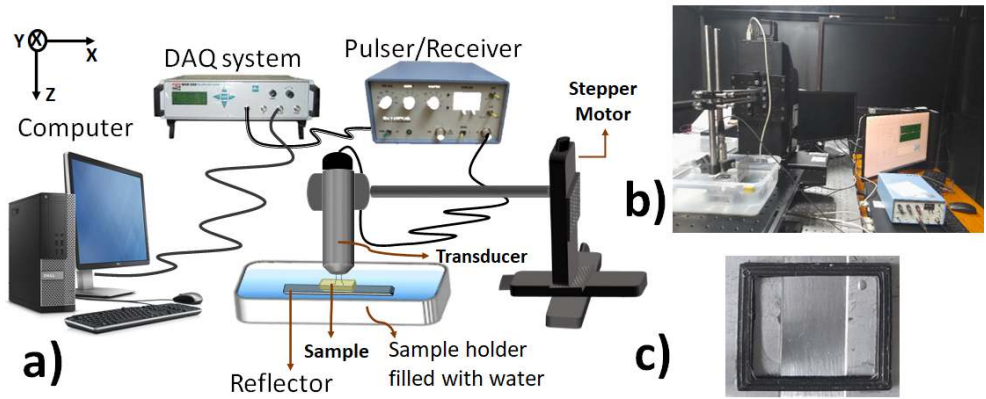

*Fig. 1 (a) A schematic representation of the experimental setup used for measuring the ultrasound velocity and attenuation coefficient in gel wax. (b) Photograph of the experimental setup. (c) Photograph of the gel wax sample (20 × 20 × 2.2 mm $^3$).*

*2.1.1 (a) Experimental procedure*: The $c_{us}$ was determined using the pulse-echo method. The US signals were obtained from the proximal and distal surfaces of the sample at times $t_1$ and $t_2$. The A-line data recorded in the experimental system for the samples are shown in Figure 2. The envelope detection of A-line data (representing the variation of reflected US signal against time) was performed using the Hilbert transform (in MATLAB), and time points corresponding to the two local maxima (refer to Figure 2 inset) determined as $t_1$ and $t_2$. Data were processed, and the $c_{us}$ was measured using the time-of-flight equation. The time interval $\Delta t$ ($\Delta t = t_2 - t_1$) was taken by the acoustic pulses in travelling to and fro, across the thickness (d) of the sample. Therefore, at time $\Delta t$, the pulses traverse the sample with an effective thickness $d_{eff}$ ($d_{eff} = 2d$), where $d_{eff} = c_{us} \times \Delta t$. The $c_{us}$ value of the gel wax was obtained using the equation $c_{us} = d_{eff}/\Delta t$. The results were an average of ten data sets.

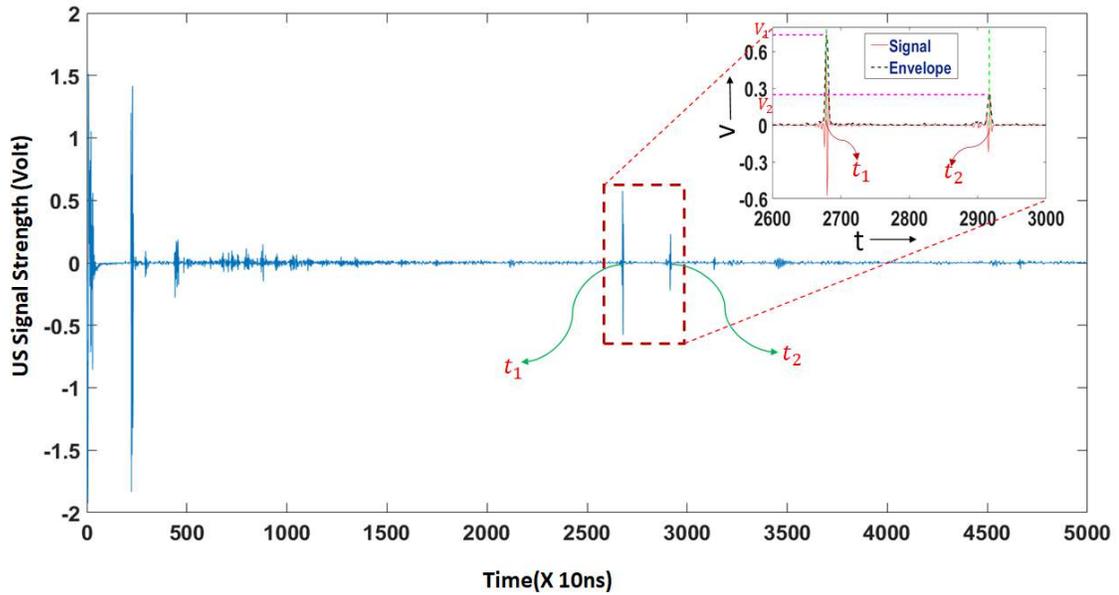

*Fig. 2 A-line graph depicting the variation in reflected ultrasound (US) signal strength over time (in nanoseconds) for the gel wax (thickness ~1.7 mm). This figure displays the signals reflected from the proximal ($V_1$) and distal ($V_2$) specimen surfaces, at times $t_1$ and $t_2$, respectively. The inset is an enlarged view within the dotted rectangular box displaying the envelope detection of the curve (shown by a red dotted curve) and two local maxima (marked by two vertical green dotted lines) to determine $t_1$ and $t_2$.*

### 2.1.2 Estimation of acoustic impedance:

The $\rho$ of the gel wax was measured. The product of the experimentally determined $c_{us}$ and $\rho$ of the gel wax yielded Z (kg/m²/s or Rayls).

### 2.1.3 Attenuation coefficient measurement

According to Beer–Lambert's law, the intensity of acoustic waves decreases exponentially while propagating through a material [17].

$$I = I_0 \times e^{-\mu \times d_{eff}}$$

Here, $I_0$ and I are the intensity of the incident wave and the intensity after transiting a distance $d_{eff}$ ($d_{eff} = 2d$) in the sample, respectively, while "$\mu$" is the attenuation coefficient.

Using the setup shown in Figure 1, μ was measured. This result was obtained using the procedure implemented in our previous study [16]. Under linear approximation, the estimated value of μ is divided by the operating frequency of the US transducer (30 MHz) to obtain the value of μ in dB/cm/MHz [17].

The AP of gel wax were compared to the reported results in soft tissue [1,14,15].

*2.2 Ultrasound phantom construction*

To determine the accuracy in distance measurement in orthogonal planes and the size of a hyperechoic target in a diagnostic US scanner, the targets were placed in homogeneous gel wax. The targets in orthogonal planes were used to determine the lateral and axial resolution. Lateral resolution describes the ability of an instrument to detect and display closely spaced objects along a line perpendicular to the major axis of the US beam. The axial resolution refers to the resolution of the equipment when the specimens lie along the beam axis.

Phantom development involved:

a) Designing and fabricating the target elements and,

b) Preparing the US phantom with the gel wax and placing the targets in the phantom to obtain the desired artifact geometry.

*2.2.1 Design and fabrication of the target elements*

The present US phantom was designed to verify the image uniformity and accuracy in distance measurements in the lateral (5 mm and 10 mm) and axial (20 mm) planes, in addition to identifying a hyperechoic target (SS-disc) embedded in the gel wax and measuring its diameter. As the test objects were immersed in a hot melted gel wax (~ 105°C) during phantom fabrication, the components should endure this environment. Therefore,

PTFE (melting point, *mp* ~ 327 °C) and SS (*mp* ~ 1400 °C) were chosen. PTFE is also suitable for US imaging [18].

PTFE sheet (20 cm × 20 cm × 5 mm), SS-disc (outer diameter = 16.8 mm), and monofilament nylon line (Thickness = 0.23 mm, Strong Force XP, Sea Rock Adventures, Mumbai, India) were procured.

Two rectangular annuli were cut from the PTFE sheet using a high-precision milling machine (Hindustan Machine Tools Ltd., Bangalore, India, Model TRM 5V), leaving 1 cm from the edges, forming two grids. The PTFE annuli were drilled using the milling machine, as per the drawings shown in Figure 3-a. Grid-1 (85 × 70 mm$^2$) contained two sets of four holes (diameter ~ 0.5 mm) in pairs at a distance of 5 mm (set-I) and 10 mm (set-II). Grid-2 (70 × 70 mm$^2$) had three pairs of holes spaced 20 mm apart (set-III). Three small sheets of PTFE (20 × 70 mm$^2$) were attached to the grids (two to Grid-1 and one to Grid-2) using SS screws. A transparent nylon line was threaded and secured with a knot in the first hole of set-I (Grid-1). This string was inserted into the opposite hole, and the next three pairs of set-I holes were woven. The SS-disc was slid at the fourth position of the grid, and the filament was sequentially threaded through the remaining holes. These formed lateral targets, separated by 5 mm and 10 mm in Grid-1. Similarly, Grid-2 was weaved to obtain the axial targets 20 mm apart. Figure 3-b shows the fabricated target elements and their alignment in the phantom.

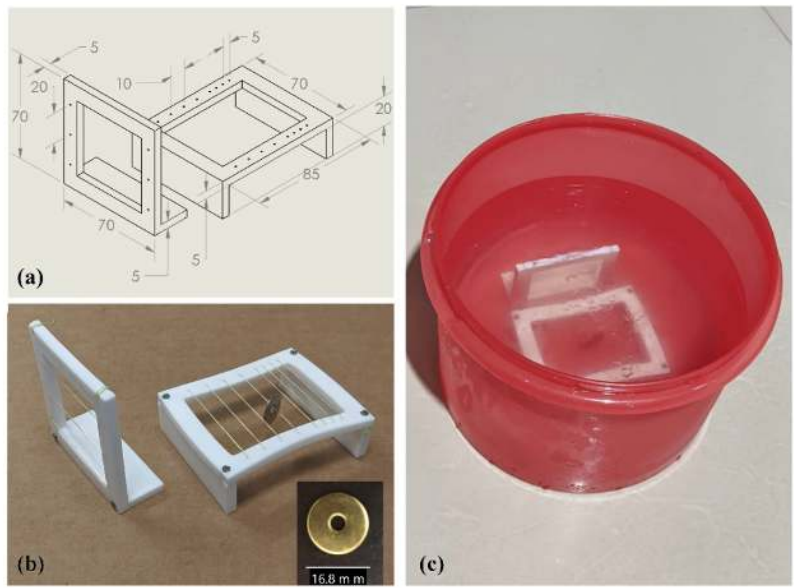

*Fig. 3 a) Drawings of the Grids-1 (Right) and 2 (Left) showing their specifications (in millimeters) to weave the nylon filaments separated by 5 mm (set-I) and 10 mm (set-II) in Grid-1 and 20 mm set-III) in Grid-2. b). The fabricated Polytetrafluoroethylene Grids-1 and 2 with the filaments and SS-disc. The diameter of the disc is shown in the inset. c) The gel wax phantom showing the grids within the transparent material.*

*2.2.2 Phantom construction*

Two *kilograms* of gel wax (Indian Wax Industries, Mumbai, India, *mp* ~ 70 °C), a SS pot (diameter = 22 cm and height = 20 cm), and an airtight container were purchased.

The gel wax was heated in the SS pot at the lowest temperature setting of an induction cooking top. After the wax melted, the temperature was raised to the next setting for 2-3 seconds and then lowered to the previous temperature. The procedure was repeated 3-5 times, and the melted wax was occasionally stirred slowly to release the dissolved air. Air bubbles prevent the transmission of US signals in the phantom body and limit their applications. To prevent air bubbles from forming on the surface of the grids, coconut oil was rubbed onto them, and excess oil was drained. Each grid was carefully placed in hot wax. Grid-1 was placed such that the filaments were parallel to the phantom surface, and in Grid-2 the plane containing the strings and phantom surface were perpendicular. This arrangement provided targets with lateral (5 and 10 mm) and axial (20 mm) geometries for distance evaluation in orthogonal planes. The cycle of maintaining the lowest temperature setting, increasing it for 2-3 seconds, and slow stirring was repeated until no air bubbles were found on visual inspection of the molten wax. The pot was then placed on a countertop and covered with a perforated paper. It is observed that slow cooling keeps air pockets to a minimum and hence, the phantom should be left in a warm environment (ambient temperature > 30 °C). After 45 min, the phantom was checked for trapped air and reheated if necessary. Air pockets were prevented by this step in the phantom prototypes. The phantom was allowed to settle overnight, and was released from the pot using a soap solution. After washing with cold water and drying, the phantom was stored in an airtight container. Figure 3-c shows the gel wax phantom with embedded grids.

*2.3 Phantom calibration and validation*

Although the targets were constructed with specific dimensions, measuring them after the high-temperature treatment was crucial. The first step was measuring and comparing the target dimensions from the CT scans and physical measurements. This will provide an estimate of the change in a target dimension following the high-temperature treatment and ensure that the composition of the gel wax is uniform without any air-pockets in the phantom. Then US imaging was performed, and deviations of US measurements from the physical dimensions were determined.

The phantom was scanned using a 16-slice CT scanner (Optima CT580 W, GE Healthcare) with exposure parameters of 120 kV, 277 mA, 24 mAs, and slice thickness of 0.625 mm. The CT images were analyzed using the Advantage Workstation for Diagnostic Imaging (ADW 2.0 CT Workstation, GE Healthcare). The uncertainty in the distance measurement of the ADW system (0.4%) was determined by measuring and comparing a specific target with its specifications from the CATPHAN phantom part # CTP503 (The Phantom Laboratories Greenwich, NY) data sheet. The results of the distance measurements of the targets in the CT images were an average of five readings for every inter-filament distance for sets I, II, and III. Similarly, the diameter of the SS-disc was determined from the CT images of Grid-1. The US images of the gel wax phantom were obtained using Samsung RS80 EVO (Samsung Medison, Gangdong, Seoul, Republic of Korea) scanner and a linear probe LA2-9A (nominal frequency 9 MHz), bandwidth 2-9 MHz, footprint 44.16 mm × 14 mm, pulse-repetition frequency 1.1 – 29.99 kHz, and gain ranging from 0-100 during imaging. Probe was placed on the phantom and the gain was adjusted to achieve the highest resolution in the target images. The gain settings were zero, 67, and 100 for resolving the targets in Set-I (see Figure 5-a and b), the SS-disc (shown in Figure 5-c), and Set-III (refer to Figure 5-d). The spacing between the nylon strings in Set-I, II and III and the diameter of the SS-disc were measured using the digital callipers of the US scanner. Each measurement in the US image was an average of five readings.

*Image uniformity:* Concentrating only on the background of the phantom image, the B-mode scan was performed by varying the gain to obtain a uniform brightness. Non-uniform images indicate the presence of weak or dead piezoelectric elements in the transducer.

*Geometric conformity and hyperechoic target visualization:* An US scan was performed to identify and measure the distance between the nylon strings in orthogonal planes. The images were improved by reducing the gain to zero. Grid-1 filaments (set-I and II) appeared as crescent-shaped speckles (refer to Figures 5-a and b) whose separation was measured. The diameter of the hyperechoic target (SS-disc) was also determined. The distance between the set-III filaments seen in Figure 5-d as continuous lines, was also obtained,.

The current phantom was validated by comparing the target measurements taken in the CT and US with the physical dimensions. The results of the CT measurements provided the specifications for the present US phantom. However, during the periodic evaluation of the current diagnostic US scanner, the results obtained from the US would be the baseline values to which the equipment should comply.

The phantom was stored in a room temperature at 25 °C and during the $1^{st}$, $10^{th}$, $30^{th}$, $54^{th}$, and $62^{nd}$ week CT scan of the phantom was performed. The CT DICOM (Digital Imaging and Communications in Medicine) images were exported to the Eclipse™ v-13.6 (Varian Medical Systems, Palo Alto, CA) system. Eclipse is a radiation therapy planning system that can calculate the distance between two points on a CT scan. The uncertainty in the distance measurements of the Eclipse$^{TM}$ system (0.2%) was obtained by measuring a target and comparing it with its specification from the CATPHAN Phantom Part # CTP504 (The Phantom Laboratories Greenwich, NY, USA) data sheet. In addition, a routine QA program based on a technical report published by the International Atomic Energy Agency, Vienna, Austria, was also conducted for the Eclipse$^{TM}$ system annually [19]. The phantom volume (cc) from the segmented CT images was obtained by auto-contouring the phantom body surface using the *Contouring* application of the Eclipse$^{TM}$ system. The change in volume over 62 weeks was estimated using this data.

## 3 Results

## 3.1 Acoustic property of the gel wax

### 3.1.1 Velocity of ultrasound

The AP values of the gel waxes are listed in Table 1. The average $c_{us}$ in the gel wax was found to be 1418.32 m/s at 20 °C. The variation in the $c_{us}$ when compared to that of soft tissue (1561 m/s) is 9.1%.

### 3.1.2 Acoustic impedance

The Z obtained from the product of $\rho$ (0.87 g/cm$^3$) and $c_{us}$ = 1418.32 m/s of the gel wax was 1.23 MRayls, as shown in Table 1.

### 3.1.3 Attenuation coefficient

The μ of the gel wax (0.88 dB/MHz/cm) was higher than that of soft tissue (Table 1).

**Table 1**: Results of the acoustic properties of the gel wax and comparison to soft tissue

| Sample | $c_{us}$(m/s) | $\rho$(g/cm$^3$) | Z (MRayls) | μ (dB/MHz/cm) |
|---|---|---|---|---|
| *Soft tissue (a) | 1561 | 1.043 | 1.63 | 0.5-0.7 |
| Gel wax (b) | 1418.32 ± 31.31 | 0.87 ± 0.02 | 1.23 ± 0.05 | 0.88 ± 0.05 |
| % variation [(b-a)/a×100] | - 9.1% | - 16.6% | -24% | Higher |

*Reference values of the velocity of ultrasound ($c_{us}$), density ($\rho$), acoustic impedance (Z), and attenuation coefficient (μ) for human soft tissues and these values in gel wax are presented. The deviation of each parameter compared to the corresponding value in soft tissue at 20 °C is also shown.*

*Source: Empirical relationships between acoustic parameters in human soft tissues; T.D. Mast (2000) derived from Duck F.A. (1990) and ICRU Report -61 [15,20,21].

### 3.2 Phantom calibration and validation

Figure 4 shows the CT images of the targets in different planes of Grid-1 (set-I, set-II, and SS-disc) and Grid-2 (set-III). The filaments at 5 and 10 mm intervals in Grid -1 are visible against the homogeneous background of the gel wax (Figure 4-a). The CT image of the phantom shows the SS-disc in Figure 4-b. Figure 4-c shows Grid-2 with filaments 20 mm apart. The arrow in the image shows the string used to secure the nylon wire, which is not a target.

*Fig. 4 Computed tomography images of the gel wax phantom*

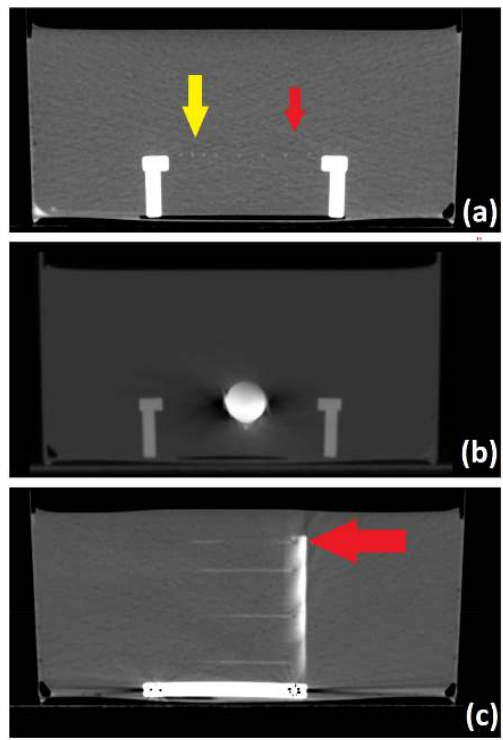

*a) Set-I (yellow arrow) and set-II (red arrow) targets at 5 mm and 10 mm distances in Grid-1 are visible against the homogeneous gel wax background. b) Image of the SS-disc in Grid-1, as seen in the CT scan. c) CT image plane showing the set-III filaments in Grid-2 separated by 20 mm; the arrow indicates the string used to secure the weaves in the grid and not a target.*
13

As the US beam is perpendicular to the plane containing the targets in Grid-1, the nylon strings appear as hyperechoic flecks on the US images (Figures 5-a and b). In Figure 5-a, the first four speckles belong to set-I (5 mm) and the remaining to set-II (10 mm). Figure 5-b shows the set-II targets, and the SS-disc can be seen in Figure 5-c. As the US beam is parallel to the plane containing the set-III filaments (Grid-2), it can be seen as a continuous line (marked with yellow arrows) in Figure 5-d.

*Fig. 5 B-mode images of the gel wax phantom*

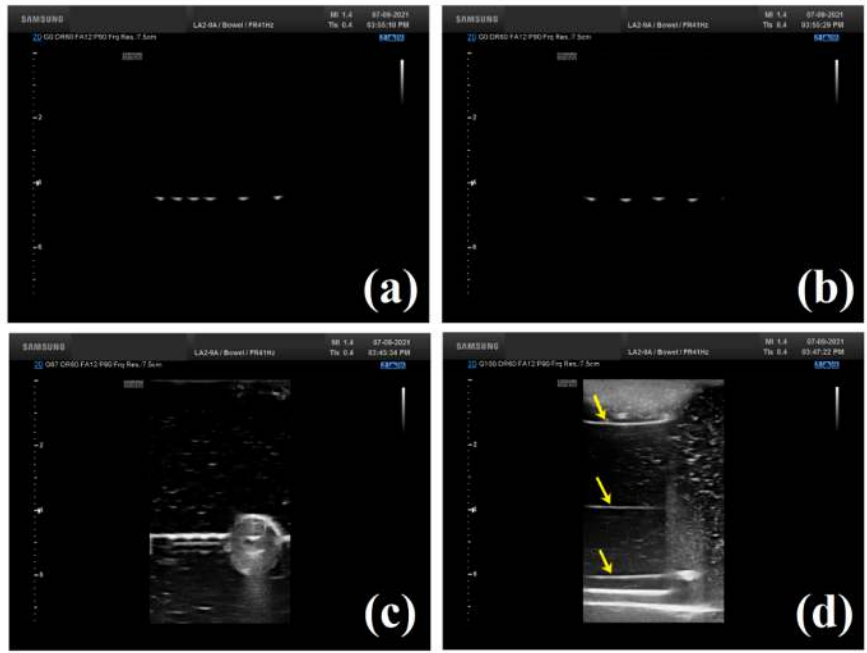

*a) The set-I (5 mm) targets in Grid-1 b) The set-II filaments separated by 10 mm in Grid-1 are seen. c) The SS-disc in Grid-1. d) Targets at 20 mm in Grid -2, the yellow arrows indicate the actual targets.*

**Table 2.** Results of the geometric conformity and hyperechoic target size measurements from the CT and US Images of the gel wax phantom

| Item | Grid-I | | | Grid-II |
|---|---|---|---|---|
| | Set-I(mm) | Set-II (mm) | SS-Disc(mm) | Set-III (mm) |
| *Actual value* (a) | 5 | 10 | 16.8 | 20 |
| *CT Image* (b) | 4.88±0.32 | 10.11±0.18 | 17.63±0.09 | 20.15±0.29 |
| *US Image* (c) | 5.26±0.30 | 10.61±0.25 | 18±0.43 | 20.88±0.13 |
| *%Variation CT [(b-a)/a×100]* | -2.4%* | 1.05% | 4.96% | 0.75% |
| *%Variation US [(c-a)/a×100]* | 5.2% | 6.1% | 7.14% | 4.4% |

*Table 2: Results obtained from the CT and US imaging of the gel wax phantom and the variation compared to the actual measurements during fabrication. * The value is smaller than the actual value.*

The mean inter-filament distances in set-I (physical distance 5 mm) are 4.88 ± 0.32 mm and 5.26 ± 0.3 mm in the CT and US images, respectively. The respective values in set-II (physical distance 10 mm) are 10.11 ± 0.18 mm and 10.61 ± 0.25 mm. The size of the SS-disc from the CT and US is 17.63 ± 0.09 mm and 18 ± 0.43 mm, respectively. The average distance in the Grid-2 targets (20 mm) is 20.15 ± 0.29 mm in CT and 20.88 ± 0.13 mm in US scans. The maximum variation between the CT and actual measurements was 4.96% when measuring the diameter of the SS-disc. However, this value for US measurements was

7.14%.

Figure 6 shows the absolute volume of the phantom 2076.22 cc, 2066.92 cc, 2048.24 cc, 2040.60 cc, and 2038.21 cc on the 1st, 10th, 30th, 54th week, and 62nd week respectively. The reduction in phantom volume from the value obtained on the first day of the measurement (2076.22 cc) and sixty-second week (2038.21 cc) was 1.83%, that is about 0.13% in a month.

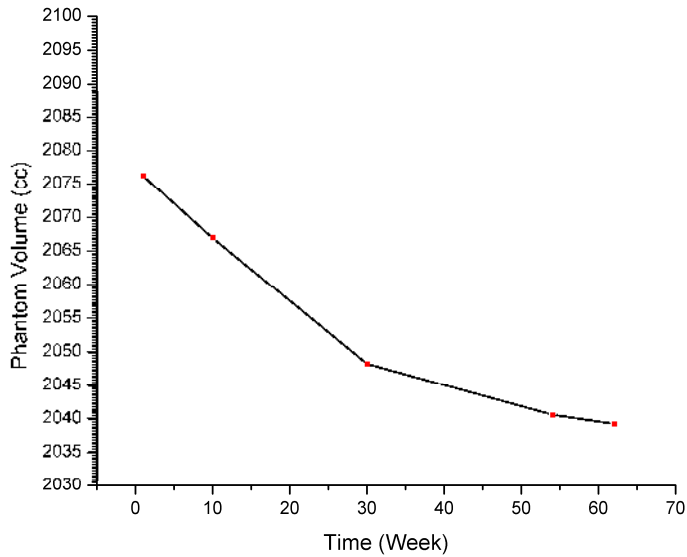

*Fig. 6 The absolute volume of the gel wax phantom measured after 1, 10, 30, 54 and 62 weeks. These data indicate a decrease of 0.13% in phantom volume over a month.*

## 4. Discussion

### *4.1 Acoustic property of the gel wax*

#### 4.1.1 Velocity of ultrasound

The measured $c_{us}$ in the gel wax (1418 m/s) is in agreement with the previously reported results 1420-1480 m/s [9,12]. *Pavan et al*. had demonstrated that the concentration of SEBS in mineral oil was directly proportional to $c_{us}$ [22]. *Oudry et al*. reported that the $c_{us}$ varied from 1420 m/s for a gel with a SEBS concentration of 2% to 1440 ms$^{-1}$ at a concentration of 6% [9]. Although the concentration of SEBS in mineral oil for the present sample of gel wax was unknown, $c_{us}$ 1418 m/s is close to 1420 m/s suggesting the concentration of the copolymer may be around 2%. When the scatterers in the scanning medium are separated by a distance smaller than the wavelength of the US wave, a weak echo appears as a complex mottled pattern (speckle). In the US examination, a TMM should reflect this pattern produced by soft tissue uniformly. This has motivated researchers to use additives such as graphite powder, micro glass spheres, other waxes, and polymers to tune the AP of materials [9,11,12,23]. The purpose of this study is to determine if it is possible to fabricate a phantom using gel wax without the use of additives. A first-hand analysis was conducted to determine whether the gel wax transmits US signals, and the embedded artifacts could be identified and resolved. As shown in Figure 5-d, the US image of gel wax does not resemble soft tissue (liver parenchyma) exactly, but it can transmit the US signal and the targets could be resolved.

*4.1.2 Acoustic impedance and attenuation coefficient*

The lower $Z$ in gel wax (refer to Table 1) when compared to soft tissue may be a reason to obtain hypoechoic background in the US images. However, this is not a limiting factor in using gel wax, as the targets could be resolved, and the subsequent measurements were satisfactory. The µ obtained here exceeds the accepted range for soft tissue, but it does not prevent us from imaging the targets.

*4.2 Phantom calibration and validation*

*Image uniformity:* The image uniformity of an US scanner is determined by scanning a homogeneous phantom while varying the gain to look for uniform brightness in the US image background. The CT images in Figure 4 show that the phantom is homogeneous, free of air pockets, and suitable for this test. In the US scan, a uniform echotexture indicates satisfactory performance from every transducer element. Defective transducers reduce the echogenic uniformity, show a gradual loss of intensity, and sometimes vertical streaks with a signal drop against a homogeneous background. [1,3]. During US imaging, the gain was set to zero so that the nylon targets are discernible in Figures 5-a and b, and the echotexture of the gel wax background is minimized. In Figure 5-c, the hyperechoic SS-disc creates a weak artifact affecting the echotexture of gel wax. Hence, Figure 5-d is used in reporting the echogenicity of the gel wax where the speckles are uniform in the region near the nylon strings. Here, a few bright speckles within the gel wax background are seen while approaching PTFE frames, caused by its higher density (2.2 g/cc) than gel wax (0.87 g/cc). However, the echotexture within the gel wax is uniform, indicating a satisfactory probe function. Any TMM for US phantoms or gel wax without a target can also be used to check image uniformity [24].

*Geometric conformity:* This test requires resolving and measuring the distance between filaments in Grids-1 and 2 present at a depth of 7.5 cm from the phantom surface with the present transducer.

Lateral Resolution: The average inter-filament distance measured in US images of the Grid-1 targets, set-I (5 mm) and set-II (10 mm), were 5.26 ± 0.3 mm and 10.61 ± 0.25 mm, respectively. The variations in these measurements were 5.2% and 6.1%, respectively, compared to the actual values in Table 1.

Axial Resolution: The average inter-filament distance from the targets of set-III (20 mm) in the US image was 20.88 ± 0.13 mm with a variation of 4.4% against the actual value, refer to Table 2.

The fuzzy appearance of the filaments in Figures 5-a and 5-b are caused by the finite resolution of the US equipment, refraction, variation of $c_{us}$ from 1540 m/s, and the presence of noise. Here, the small point targets (nylon filaments) in the grids appeared as short crescent-shaped dots making it challenging to accurately localize the actual point of interaction and influences the target measurement in the US images [25]. It was easier to distinguish the filaments using a 9 MHz transducer than a 3 MHz, proving that the transducer resolution increases with frequency. In Grid-2 the finite beamwidth of the transducer prevents viewing the entire filament in one image. Unlike Grid-1 targets, the thickness of the filaments in this grid remains unchanged in the US image as the inclination of the scanning plane relative to the beam direction is significantly less [26].

*Hyperechoic target visualization:* The metal specimen was critical as it could retain its shape post-heating and produce different image contrast than the nylon strings. An US image of a metal specimen usually shows only the section facing the US beam owing to partial transmission. However, the present specimen with the thickness (0.02 mm) and volume (0.004 cc) was chosen prudently, which was insufficient to attenuate the US signal significantly, and as seen in Figure 5-c, the disc was visible during scanning. The diameter of the SS-disc varied by 5% and 7.1% in CT (17.63 ± 0.09 mm) and US (18 ± 0.43 mm) images, respectively, compared to its physical size (16.8 mm). Among all the measurements, this result deviated the most from its actual value.

The distance measurements from the CT results in sets I, II, and III varied by 2.4%, 1%, and 0.8%, respectively, compared to the actual values of these targets. Because SS retains its shape post-heating at 105°C, the difference in measuring the diameter of the SS-disc (5%) is caused by artifacts in the CT image. Throughout the heating and cooling cycle, all targets remain within 5% of their original geometry.

As reported in a rigid flow study using doppler, PTFE has a $c_{us}$ 1376 ± 40 m/s and a high melting point [27]. These characteristics make PTFE preferable for similar applications. The phantom images also show that the nylon filaments retained their shape after heating. The SS-disc was used to simulate a hyperechoic target but a material with fewer artifacts should be preferred. Researchers have used graphite powder ($mp$ = 3600 °C) mixed with other materials like agar to create targets with varying echogenicity [23,28]. But with their irregular geometry the geometric accuracy of the targets cannot be verified. Different CIRS (Computerized Imaging Reference Systems Inc., Norfolk, VA, USA) phantoms have been investigated for the ability of the US equipment to measure the $c_{us}$ in targets with varying densities (echogenicity) and to develop indigenous scanners [29,30]. Phantoms with targets in various geometries and wires in different configurations have been investigated earlier to assess the resolution in US scanners [31–35]. QA in these wire phantoms required solving a mathematical problem to obtain the phantom coordinate system from the US image coordinate system of the targets using a transformation matrix. Here, point reconstruction accuracy was the parameter that decided a more robust system. Some phantoms had wires crossing over, which required identifying the actual targets in the scans. Data from several scan images were acquired from various positions and manually segmented to solve the problem. The accuracy of such equipment was high because they were meant for surgical procedures. However, the present phantom is designed for US scanners used in clinical diagnosis which is possible with an uncertainty of 7% because to measure a distance of 10 mm, the callipers will display 10 ± 0.7 mm, which is acceptable. The targets of this phantom have a simple geometry, and the scans can be performed easily without any ambiguity. A probe is moved over a phantom to obtain images for analysis and does not require processing or technical expertise.

Over 62 weeks there is no change in structural stability, phantom volume, or discoloration in the phantom. The transmission of US in the phantom (echogenicity), relative positions and the geometry of the targets was also maintained during this period. Materials like agar deteriorate by microbial infestation and evaporation, unlike oil. According to the manufacturer, gel waxes last longer when stored at ambient temperature in an airtight container. A long shelf life, preferably 5-10 years, is desired for a QA phantom for diagnostic US scanners, and oil-based gels can satisfy this requirement. Considering these properties of gel wax, a change of this magnitude (1.8% in 14 months) is reasonable and meets the stability requirement of the phantom for long-term use.

Melted gel wax produces fumes, so it is advisable to use a mask and work in a well-ventilated area. Moreover, low and high frequency probes have different phantom requirements. Therefore, it is practically difficult to carry out these checks with a single phantom. A possible solution can be to design targets near the phantom surface at < 5 cm depth and > 7 cm deep to perform shallow and deep imaging.

The average $c_{us}$ measured in the gel wax 1418 m/s was 9% lower when compared to soft tissue and is not a major restriction in similar applications, as the targets are resolved within 7% accuracy. The transparent gel wax enables removing air pockets during phantom preparation by visual inspection. The phantom target geometry allows measuring multiple distances (5 mm, 10 mm, and 20 mm) in both lateral and axial directions. The embedded SS-disc resembled a target with a high reflection coefficient compared to gel wax. Gel wax is an affordable system that can be molded into any desired shape with customizable targets based on individual requirements even without additives to design a reliable phantom.

According to these results, the current phantom can be used in routine QA of diagnostic US scanners to measure distance with an accuracy of 7% and image uniformity in the frequency range of 2–9 MHz. The distance between the targets in set-I (5.26 ± 0.3 mm), set-II (10.61 ± 0.25 mm), set-III (20.88 ± 0.13 mm), and the diameter of the SS-disc (18 ± 0.43 mm) obtained from the US measurements will be the baseline values of the present US scanner. However, phantom specifications can be recorded based on the results of CT imaging.

## 5 Conclusions

This study demonstrates the development of an US QA phantom using gel wax. The AP of gel wax was experimentally determined, which was consistent with the existing literature. In validation testing, the phantom was accurate within 7% in measuring distance, and its volume remained stable within 2% over a year. This portable phantom can be of interest in periodic assessment of US scanners across multiple centers because of its structural stability, low cost, and customizable design.

**Funding**

This research did not receive any specific grants from funding agencies in the public, commercial, or not-for-profit sectors.

**Declarations**

**Conflict of interest:**

The authors declare that they have no conflict of interest.

**Ethical approval:**

This article does not contain any studies performed on human participants or animals.

[26] Dendy PP & Heaton, B. Physics for Diagnostic Radiology (3rd ed.). 3rd ed. CRC Press.; 2011.

[27] Wong EY, Thorne ML, Nikolov HN, Poepping TL, Holdsworth DW. Doppler Ultrasound Compatible Plastic Material for Use in Rigid Flow Models. Ultrasound Med Biol 2008;34:1846–56. https://doi.org/10.1016/j.ultrasmedbio.2008.01.002.

[28] De Carvalho IM, Basto RLQ, Infantosi AFC, Von Krüger MA, Pereira WCA. Breast ultrasound imaging phantom to mimic malign lesion characteristics. Phys. Procedia, vol. 3, Elsevier; 2010, p. 421–6. https://doi.org/10.1016/j.phpro.2010.01.055.

[29] Taylor Z, Jonveaux L, Caskey C. Development of a Portable and Inexpensive Ultrasound Imaging Device for Use in the Developing World. URL Httpswww Semanticscholar OrgpaperDevelopment---Portable--Inexpensive-Imaging-Taylor-Jonveaux4a45c73373a1e1168f7a276d977d65cebc29e0eb 2017.

[30] Khalitov RS, Gurbatov SN, Demin IY. The use of the Verasonics ultrasound system to measure shear wave velocities in CIRS phantoms. Phys Wave Phenom 2016;24:73–6. https://doi.org/10.3103/S1541308X16010143.

[31] Ali A, Logeswaran R. A visual probe localization and calibration system for cost-effective computer-aided 3D ultrasound. Comput Biol Med 2007;37:1141–7. https://doi.org/10.1016/j.compbiomed.2006.10.003.

[32] Soehl M, Walsh R, Rankin A, Lasso A, Fichtinger G. Tracked ultrasound calibration studies with a phantom made of LEGO bricks. Med. Imaging 2014 Image-Guid. Proced. Robot. Interv. Model., vol. 9036, SPIE; 2014, p. 90362R-90362R. https://doi.org/10.1117/12.2044121.
25